# Ferromagnetic ordering in dilute magnetic dielectrics with and without free carriers

## K. Kikoin

*School of Physics abd Astronomy, Tedl-Aviv Universiy, Tel-Aviv, 69978, Israel*

**Abstract**

The state of art in the theoretical and experimental studies of transition metal doped oxides (dilute magnetic dielectrics) is reviewed. The available data show that the generic non-equilibrium state of oxide films doped with magnetic impurities may either favor ferromagnetism with high Curie temperature or result in highly inhomogeneous state without long-range magnetic order. In both case concomitant defects (vacancies, interstitial ions play crucial part.

## 1. Introduction

The available experimental information and existing theoretical considerations about the nature of ferromagnetic ordering in dilute magnetic dielectrics (DMD) (mostly, oxides doped with transition metal ions) [1] clearly demonstrate essential differences between these wide-gap materials and dilute magnetic semiconductors (DMS) [2]. In this paper we summarize the inherent features of DMD, which allow treating these materials as a separate family of dilute magnetic materials. In the most of DMD with high Curie temperature $T_C$, the room-temperature ferromagnetism (FM) is apparently not related to high concentration of free carriers [1,3]. This implies that the carrier-mediated exchange mechanism responsible for FM order in DMS [2,4] is ineffective in DMD. These materials are characterized by extreme sensitivity to the growth and annealing conditions [5], and practically in all cases one may conclude that the indirect exchange between magnetic transition metal ions is mediated by complex defects involving oxygen vacancies, shallow impurities or other imperfections [1,3,5,6].

We review the experimental and theoretical situation with materials, in which the role of magnetic inclusions is ruled out (II-VI oxides and some dioxides diluted with impurities of iron group) and discuss in some details possible mechanisms of indirect ferromagnetic exchange specific for the wide gap materials. Among possible complex defects, which may mediate the indirect ferromagnetic exchange between transition metal (TM) impurities in oxides, one may mention oxygen vacancies (both isolated and bound with impurities), defect magnetic polarons and bound excitons.

**2. Basic Experimental Facts**

It is generally recognized in current literature that the structural and magnetic properties of TM-doped oxide films are extremely sensitive to sample preparation and thermal processing methods [5,6]. Available methods (ion implantation, pulsed laser deposition, reactive magnetron sputtering, etc) produce imperfect films, which are far from thermodynamic equilibrium. These materials are unstable against various heterogeneities (see, e.g., [7]), namely, precipitation of other crystallographic phases, phase separation in host material, spinodal decomposition of dopant, diffusion and implantation profiles, etc. Even in carefully checked conditions, where the precipitation of parasitic phases and aggregation of superparamagnetic clusters with excessive concentration of TM ions are prevented or at least controlled, one cannot get rid of this generic feature of oxide DMD materials. We have chosen for our review two families of DMD, namely ZnO and $TiO_2$ doped with iron group ions (V, Cr, Mn, Fe, Co). The reason for this choice is a consensus within the experimental community about intrinsic nature of long range FM order in these materials, at least in the best samples, although the notion "best" demands additional specification, which will be given below.

One of the most salient features of TM doped oxide films is stabilization of FM order at room temperature already in relatively weakly doped *n*-type materials with concentration *x* of magnetic impurities well below the percolation threshold $x_c$ for disordered magnetism [1]. To resolve this discrepancy, the authors of the review [1] proposed the concept of weakly bound magnetic polarons: these are the donor electrons near the bottom of conduction band, which are captured into a spin-split impurity band by means of *s-d* exchange with TM impurities. The radius of polaron orbitals is large enough to provide the indirect exchange between magnetic ions via the states in the partially filled polaronic band even at $x < x_c$ . This approach may explain appearance of FM order in n-type samples with metallic-type conductivity like (Zn,Mn)O and (Zn,Co)O codoped with Al [8]. However, FM order is observed also in highly insulating samples of oxygen deficient TM doped ZnO [3,9] and TiO$_2$ [5,10,11], so the concept of magnetic polaron should be modified in order to interpret these data. The most exhaustive data about dependence of FM order on the degree of imperfection and codoping level in n-type ZnO doped with Ti, V, Mn and Co are published in [3]. In the experiment both oxygen pressure and codoping level were varied. Thus, the wide range of samples from strongly non-stoichiometric oxygen-deficient films to overcompensated n-type metallic-type films was investigated. The data for (Zn,Co)) codoped with Al are presented in Fig.1. Three different transport and magnetization regimes were identified. In the insulating regime (1) of low carrier concentration the films are magnetic at room temperature, and the electron transport arises from a variable range hopping, where the resistivity obeys Mott's law, $\rho(T) = A \exp(T_0/T)^{1/4}$ . The samples are strongly magnetic only if $T_0 > 10^4$ K, i.e. if hopping kinetic energy is small enough. In the intermediate regime (2) magnetization disappears, but with growing donor concentration it arises again and magnetization in metallic phase (3) has a sharp maximum as a function of carrier density. Evolution of magnetization and the carrier density with decreasing temperature from 295 K to 5 K for three samples a,b,c is also shown. Similar effect, i.e. existence of FM order both in insulating and metallic phases was observed in (Ti,Co)O$_2$ films [5,12]. Apparently, existence of magnetism both in metallic and highly resistive films is the generic property of dilute magnetic oxides, which should be explained within a framework of microscopic theory.

Although the most of concise experimental data are related to *n*-type samples, one should mention the peculiar observation of polarity dependent ferromagnetism in TM doped ZnO [13]. It was shown that ferromagnetism in quite perfect (Zn,Co)O and (Zn,Mn)O nanocrystalline films prepared by direct chemical synthesis, have opposite carrier polarities. FM order is observed in *p*-type (Zn,Mn)O and in *n*-type (Zn,Co)O. Changing the type of polarity by means of nitrogen codoping from *p*- to *n*-type in the first case and from *n*- to *p*-type in the second case resulted in disappearance of the FM order. This remarkable result is worth of attention because these chemically prepared samples are relatively free from structural defects and inclusions.

**3. Basic Theoretical concepts**

To interpret the above experimental data, one should have in mind the generic inhomogeneity of thin films containing both intrinsic defects (oxygen vacancies), TM ions in concentration well above the solubility limit, and uncontrollable concomitant defects. Additional source of inhomogeneity is the film surface [6]. Any thermal treatment induces the diffusion of vacancies, magnetic ions and other defects and formation of complexes. Even in the samples, where the spinodal decomposition and accumulation of defects near the film surface do not make the film heterogeneous, and the FM order is formed in a bulk of the sample, one may be sure that the complex defects are responsible for this order rather than isolated TM ions. Such picture is accepted in many recent theoretical models [14-17].

The complex defects involving magnetic ions in substitution or interstitial positions, oxygen vacancies and adjacent ions from host cation sublattice are formed in the course of migration of these defects under thermal processing. An example of such complex is the double defect [$Co_{Ti}$,$V_O$] in (Ti,Co)$O_2$ discussed in [14]. When Co impurity substitutes for a Ti ion in the vicinity of oxygen vacancy $V_O$, the bound state is formed in accordance with "reaction"

$$Co^{4+}(d^5) + V_O(p^2) \rightarrow Co^{2+}(d^7) + V_O(p^0) \qquad (1)$$

In accordance with this reaction two electrons from double donor $p$-orbital level $\varepsilon_t$ of O vacancy in the upper part of the forbidden gap of TiO$_2$ move to the $3d$ shell of substitution Co ion, to the energy level determined by the addition energy [18,19]

$$\varepsilon^{(n/n-1)} = E\ (d^n) - E(d^{n-1}) - \ \varepsilon_t \qquad (2)$$

($n$=7) in the lower part of the energy gap. The binding energy of the complex [$Co_{Ti}$,$V_O$] is determined by the charge transfer energy $\varepsilon_t$ - $\varepsilon^{(n/n-1)}$ and the energy of Coulomb attraction between negatively charged impurity and positively charged vacancy. Such situation is common for TM-doped oxides, because of the universal trend in the position of addition energies (2) relative the center of gravity of the continuum of host valence and conduction bands.

In II-VI oxides the levels $\varepsilon^{(n/n-1)}$ corresponding to the neutral substitution position of TM ion in the cation sublattice form the mid-gap states between the top of the valence band $\varepsilon_t$ and the bottom of conduction band $\varepsilon_b$ [19-21]. Neutral $Mn^{2+}(d^5)$ is the only exclusion: its $\varepsilon^{5/4}$ d-level D(0/+) falls deep in the valence band of ZnO [21]. Corresponding universal trend for TM-doped $TiO_2$ is more complicated [19,22]: several charged donor states D(-/0), D(--/-) above $\varepsilon_t$ may arise for each TM impurity. As a result, different TM ions may enter the impurity-vacancy complex in different charge states, $TM^{2+}$ or $TM^{3+}$ [23].

Eventually the thermally treated film may be considered as a non-equilibrium ensemble of oxygen vacancies partially free and partially bound with magnetic ions in a form of complexes. As is known [16], $V_O$ orbitals form a band of relatively shallow defects states slightly below $c_b$ . If the distribution of vacancies and magnetic ions is homogeneous enough after thermal treatment, then the overlap between the weekly bound $p$-orbitals of oxygen vacancies play the same part as the donor states in the polaronic model of Ref. [1]. In the films codoped with other shallow donors [3,8], second polaronic band of donor origin appears. The resulting energy level scheme is presented in Fig. 2. Evolution of this spectrum corresponding to effective carrier concentration shown in Fig.1, is presented at the upper panel. In undoped samples double donors tates $V_O(p^2)$ are partially compensated by acceptor impurities $Co_{Ti}$ in accordance with (1), so that the $V_O$-related band is partially filled and magnetized due to the indirect exchange between magnetic impurities [14]. Donor impurities compensate $[V_O-Co_{Ti}]$ complexes. In the intermediate phase the vacancy band is filled and the donor band is nearly empty. With further increase of donor concentration this band gradually fills up, and magnetizes respectively in accordance with the polaronic mechanism [1]. When the band filling process is completed, the film returns into nonmagnetic state. Evolution of magnetization is shown in the lower panel of Fig. 2.

Similar situation should be realized in TM doped ZnO with one important reservation. $TM_{Zn}$ substitution impurities are isoelectronic acceptors $TM^{2+}$, so the charge transfer from $V_O$ states is possible only provided the energy level $TM^{+/2+}$ is below or at least in resonance with $V_O$ related band states. This is the case of Cr and Co ions [10], and just for these impurities the dependence $M(n_c)$ observed experimentally [3] corresponds to the qualitative picture of Fig. 2. Microscopic calculation of this dependence may be done within a framework of generalized Alexander-Anderson model [14] with the Hamiltonian

$$H = \sum_j H_{mj} + \sum_{b=c,v} H_b + \sum_{d=\kappa,s} H_d + H_{hyb} \qquad (3)$$

This Hamiltonian includes impurity terms $H_{mj}$ describing TM ions in the sites $j$, band Hamiltonians $H_b$ with terms describing mobile electrons in the valence $(b=v)$ and conduction $(b=c)$ bands, defect Hamiltonian describing the $V_O$ related band $(d=\kappa)$ of weakly localized electrons and the set of shallow donor states $(d=s)$ near the bottom of conduction band. The last term describes the hybridization between $3d$ of TM impurities and all band states, both conducting and localized. If the concentration of TM ions is below the percolation limit for direct exchange and indirect exchange via valence and conduction bands, then its single-site part may be taken into account by means of exact canonical transformation [24]. Then the superexchange via empty states in the bands $H_c$ which has the FM character [14], is responsible for long-range order in homogeneous films. The Curie temperature is proportional to the strength of the effective inter-impurity coupling, which has the form $J_{jl}F(\mu)$. Here the constant

$$J_{jl} \sim |V_j|^2 \, |V_l|^2 \, / \rho_c |\Delta_{dc}|^2 \qquad\qquad\qquad (4)$$

contains parameters of hybridization $V_{j(l)}$ between the TM impurity in the site $j(l)$, and the states in the defect band $c$ with average density of states $\rho_c$ and the charge transfer energy $\Delta_{dc}$ between defect level and the level $E^{n+1/n}$. The function $F(\mu)$ depends on the position of chemical potential $\mu$ in the defect band which is regulated by the occupation degree. This function with a maximum around the half-filling of the defect band is shown in Fig. 3. It correlates with experimentally found variation of magnetization as a function of the carrier deficiency [3,8] (see the upper panel of Fig.1 for comparison). This trend reflects the obvious fact that the maximum energy gain due to indirect exchange via empty states in the mediating subsystem is achieved when the band of these states is half-filled. Similar trend was predicted for the indirect exchange via an impurity band split from the valence band due to magnetic impurity scattering in DMS [25].

The explanation of the dependence of magnetic ordering on the carrier polarity in nanocrystalline chemically synthesized films [13] is based on the idea of indirect exchange via bound exciton states. Such states in TM doped II-VI materials were discovered experimentally and discussed theoretically in 80-es [26,27]. An electron-hole pair may be captured by a neutral magnetic impurity in two ways in accordance with "reactions"

$$d^n + [e,h] = [d^{n+1}h] \qquad\qquad\qquad (5a)$$

or

$$d^n + [e,h] = [d^{n-1}e], \qquad\qquad\qquad (5b)$$

where either electron or hole is captured in the unfilled 3d shell of TM ion, whereas the second carrier (hole or electron) is retained on the loosely bound hydrogen-like orbit by the attractive Coulomb force. Then the indirect exchange between adjacent TM ions via virtual excitation of such electron-hole pairs namely,

$$d_i^n d_j^n \rightarrow [d_i^{n-1} e_\uparrow] d_j^n \rightarrow d_i^n [d_j^{n-1} e_\uparrow] \rightarrow d_i^n d_j^n, \qquad (6a)$$

$$d_i^n d_j^n \rightarrow [d_i^{n+1} h_\downarrow] d_j^n \rightarrow d_i^n [d_j^{n+1} h_\downarrow] \rightarrow d_i^n d_j^n, \qquad (6b)$$

is possible. The first of these processes is realized in $p$-type (Zn,Mn)O with $n$=5, and the second one may be realized in $n$-type (Zn,Co)O with $n$=7. The loosely bound hydrogen-like charge carriers play the same part in the excitonic mechanism of indirect exchange as the vacancy-related carriers in (4) or spin-polaron states in the mechanism proposed in [1]. It is worth to emphasize that the Hund rule, which is responsible for the high spin state of the 3d-electrons in the configurations $d^{\,n}$ and $d^{\,n-1}$ plays the decisive part in the ferromagnetic type of indirect exchange in these materials [14].

**Concluding remarks**

The main message conveyed in this review is that the ferromagnetic ordering in TM doped oxide films is intimately connected with the non-equilibrium state of these materials. Long range order may arise due to balance between the distribution of subsitution magnetic impurities and native defects like vacancies, interstitial impurities ets. Great scatter of magnetic characteristics is an unavoidable property of DMD. In many samples the inhomogeneous distribution of all components in this ensemble is detrimental for the FM ordering. Imperfections responsible for formation of loosely bound defect states in the energy gap of host material mediate an indirect exchange between TM impurities, and the careful control of these defects is a key to stable FM ordering with high Curie temperature in DMD.

Author is indebted to V. Fleurov and A. Rogalev for valuable discussions.

# References

[1] J.M.D. Coey, M. Venkatesan, and C.B. Fitzgerald, Nat. Mater. 4 (2005) 173

[2] T. Jungwirth, et al., Rev. Mod. Phys. 78 (2006) 809

[3] AJ. Behan, et al, Phys. Rev. Lett. 100 (2008) 047206

[4] T. Dietl, Nat. Mater. 5 (2006) 673

[5] K.A. Griffin, et al, Phys. Rev. Lett. 94 (2005) 157204

[6] B.K. Roberts, et al, Appl. Phys. Lett. 92 (2008) 162511

[7] T. Dietl, J. Appl. Phys. 103 (2008) 063918

[8] X.H. Xu et al, J. Appl. Phys. 101 (2007) 07D111

[9] C.B. Fitzgerald et al, Appl. Surf. Sci. 247 (2005) 493

[10] T. Dietl, Semicond. Sci. Technol. 17 (2002) 377

[11] T.C. Kaspar et al, Phys. Rev. Lett. 95 (2005) 217203

[12] T. Fukumura, et al, New J. Phys. 10 (2008) 055018

[13] K.R. Kittilstved, et al, Nat. Mater. 5 (2006) 291

[14] K. Kikoin and V. Fleurov, Phys. Rev B 74 (2006) 174407

[15] H. Weng, et al, Phys. Rev. B 69 (2004) 125219

[16] V.I. Anisimov, et al, J. Phys.: Cond. Mat. 18 (2006) 1695

[17] A.L. Rosa and R. Ahuja, J. Phys.: Cond. Mat. 19 (2007) 386232

[18] A. Zunger, *in* Solid State Physics, eds.D. Turnbull and H. Ehrenreich,
    (Academic Press, Orlando, 1986), Vol. 39, p. 276

[19] K. Kikoin and V. Fleurov, Transition Metal Impurities in Semiconductors,
    World Sci. (1994)

[20] V.I. Sokolov, Sov. Phys. – Solid State 29 (1987) 1061

[21] T. Dietl, Semicond. Sci. Technol. 17 (2002) 377

[22] K. Mizushima, M. Tanaka, and S. Iida, J. Phys. Soc. Jpn (1972) 1519

[23] K. Kikoin and V. Fleurov, J. Magn. Magn. Mat. (2007) 2097

[24] G. Cohen, V. Fleurov, and K. Kikoin, J. Appl. Phys., 101 (2007) 09H106

[25] A. Chattopadhyay, S. Das Sarma, and A.J.Millis, Phys. Rev. Lett. 87 (2001) 227202

[26] V.N.Fleurov and K.A.Kikoin, Solid State Commun. 42 (1982) 353

[27] V.I. Sokolov and K.A. Kikoin, Sov. Sci. Rev. A 12 (1989) 147

**Figure captions**

Fig. 1. Experimental dependence of the room temperature magnetization on the carrier density in (Zn,Co)O [3].

Fig. 2. Upper panel: evolution of energy spectrum of oxygen deficient $(Ti,Co)O_2$ (n=7) with changing degree of disorder. From left to right: insulating phase with partially filled $V_O$ related band; intermediate phase with filled vacancy band; n-doped metallic phase with partially occupied donor impurity band; metallic phase with filled donor band. Lower panel: Change of Curie temperature from insulating phase to intermediate and metallic phases.

Fig. 3. Theoretical dependence of Curie temperature on the filling of defect-related band.

Fig. 1

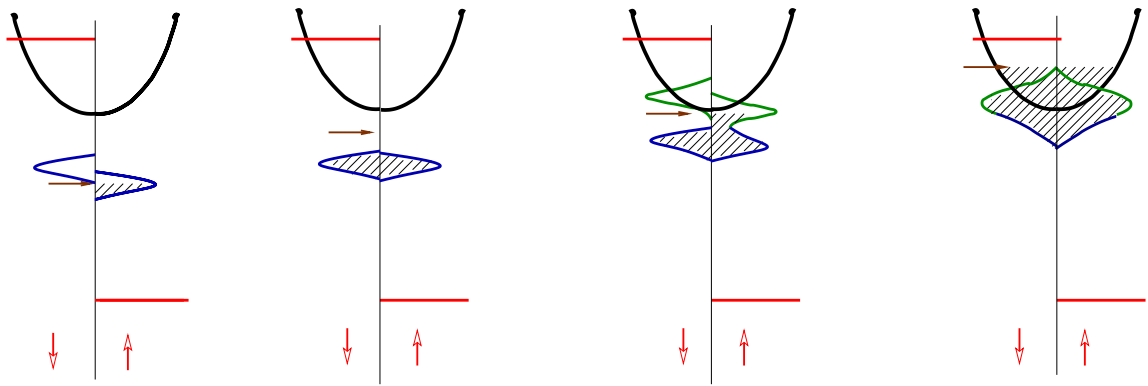

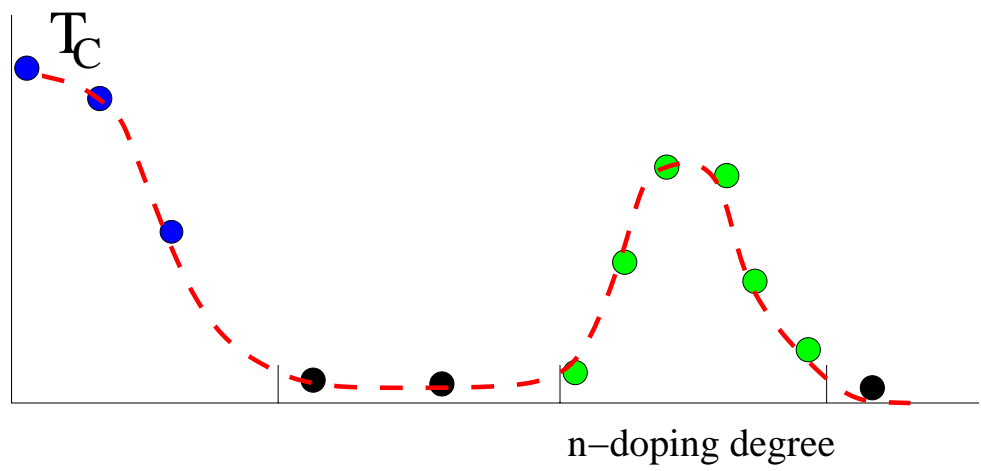

Fig. 2

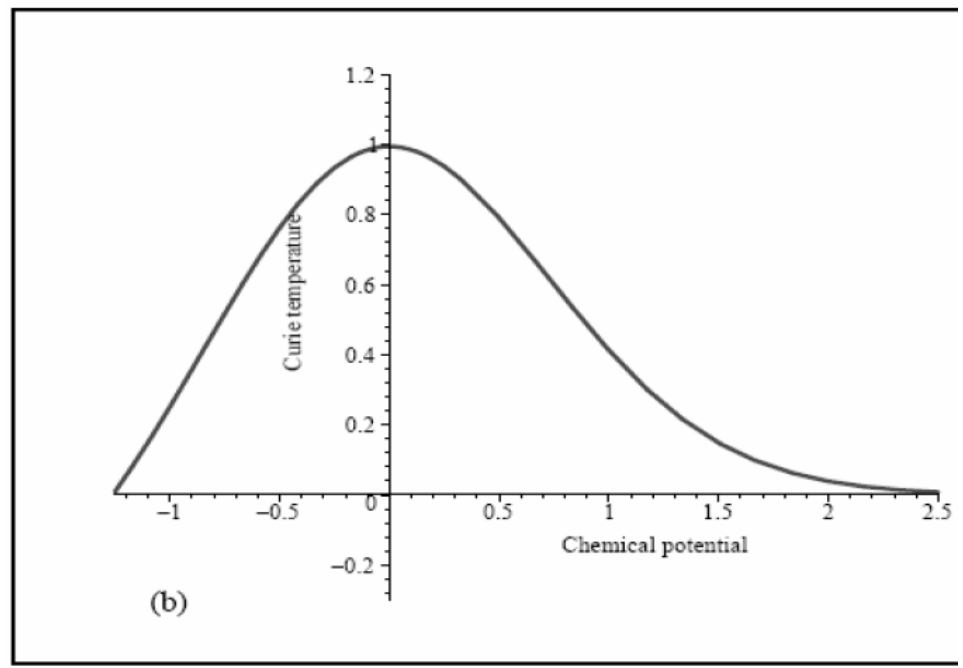

Fig.3